\documentclass[letterpaper,journal,twoside]{IEEEtran}
\usepackage{url}
\usepackage[sort,nocompress]{cite}
\usepackage{amsmath,amssymb,amsfonts}
\usepackage{xcolor}
\usepackage[ruled,linesnumbered,vlined]{algorithm2e}
\usepackage{array}
\usepackage{textcomp}
\usepackage{stfloats}
\usepackage{verbatim}
\usepackage{graphicx}
\usepackage{bbm}
\usepackage{multirow}
\usepackage{leftindex}
\usepackage{latexsym}
\usepackage{mathrsfs}
\usepackage{setspace}
\usepackage{booktabs}
\usepackage{color}
\usepackage{comment}
\usepackage[font=footnotesize,labelsep=period]{caption}
\usepackage{makecell}
\usepackage{threeparttable}
\usepackage{colortbl}
\usepackage{float}
\usepackage{lipsum}
\usepackage{bm}
\usepackage{balance}
\usepackage{cases}

\usepackage{multirow}
\usepackage[subrefformat=parens,justification=centering]{subcaption}
\makeatletter
\renewcommand{\maketag@@@}[1]{\hbox{\m@th\normalsize\normalfont#1}}%
\makeatother
\newcolumntype{P}[1]{>{\centering\arraybackslash}m{#1}}
\newcommand{\T}{\mathsf{T}}
\newcommand{\m}{\,\mathrm{m}}

\newcommand{\mm}{\,\mathrm{mm}}
\newcommand{\cm}{\,\mathrm{cm}}
\newcommand{\cw}{_\mathrm{c}^\mathrm{w}}

\newcommand{\rom}[1]{(\romannumeral #1)}

\usepackage{flushend}
\usepackage[colorlinks=true,allcolors=blue,urlcolor=blue]{hyperref}

\let\oldeqref\eqref
\renewcommand{\eqref}[1]{\textcolor{blue}{\oldeqref{#1}}}
\definecolor{mBlue}{RGB}{145,39,153}

\begin{document}
% \begin{CJK}{UTF8}{gbsn}

\title{\LARGE Passive Multi-Target Visible Light Positioning Based on\\Multi-Camera Joint Optimization
}%

\author{Wenxuan~Pan,~Yang~Yang,~\IEEEmembership{Senior~Member,~IEEE},~Dong~Wei,~Meng~Zhang,~and~Zhiyu~Zhu%
    \thanks{\textcolor{mBlue}{This manuscript has been accepted for publication in the \textsc{IEEE Communications Letters} [DOI: \textcolor{blue}{\href{https://dx.doi.org/10.1109/LCOMM.2025.3598352}{10.1109/LCOMM.2025.3598352}}]. The final version will be available at the IEEE \textit{Xplore}\textsuperscript{\textregistered}.}}%
    \thanks{This work was supported in part by National Natural Science Foundation of China under Grant 62371065 and Grant 61871047, in part by Beijing Natural Science Foundation under Grant L222043, in part by National Key Research and Development Program of China under Grant 2023YFB2904201, and in part by BUPT Innovation and Entrepreneurship Support Program under Grant 2025-YC-S007. \textit{(Corresponding author: Yang Yang.)}}%
    \thanks{Wenxuan Pan and Yang Yang are with Beijing Key Laboratory of Network System Architecture and Convergence, School of Information and Communication Engineering, Beijing University of Posts and Telecommunications, Beijing 100876, China (e-mail: \mbox{pwx@bupt.edu.cn}; \mbox{yangyang01@bupt.edu.cn}).}%
    \thanks{Dong Wei and Meng Zhang are with the Institute of Information Engineering, Chinese Academy of Sciences, Beijing 100085, China (e-mail: \mbox{weidong@iie.ac.cn}; \mbox{zhangmeng@iie.ac.cn}).}%
    \thanks{Zhiyu Zhu is with College of Physics and Electronic Engineering, Shanxi University, Taiyuan, Shanxi 030006, China (e-mail: \mbox{zhiyu.zhu@sxu.edu.cn}).}%
    \vspace{-0.7cm}%
}

\markboth{\LaTeX\ Class Files Letters,~Vol.~14, No.~8, June~2025}%
{Shell \MakeLowercase{et al.}: A Sample Article Using IEEEtran.cls for IEEE Journals}

\IEEEpubid{
    \parbox{\textwidth}{\centering 0000-0000~\copyright~2025 IEEE.}
}
% Remember, if you use this you must call \IEEEpubidadjcol in the second column for its text to clear the IEEEpubid mark.

\maketitle

\begin{abstract}
Camera-based visible light positioning (VLP) has emerged as a promising indoor positioning technique. However, the need for dedicated luminaire infrastructure and on-target cameras in existing algorithms may limit their scalability and increase deployment costs. To address these limitations, this letter proposes a passive VLP algorithm based on Multi-Camera Joint Optimization (MCJO). In the considered system, multiple ceiling-mounted pre-calibrated cameras continuously capture images of targets with unmodulated point light sources, and can simultaneously localize these targets at the server. In particular, MCJO comprises two stages: It first estimates target positions via linear least squares (LLS) from multi-view projection rays; then refines these positions through nonlinear joint optimization to minimize the reprojection error. Simulation results show that MCJO can achieve millimeter-level accuracy, with an improvement of 19\% over an LLS-based state-of-the-art algorithm. Experimental results further show that MCJO achieves an average position error as low as 5.63 mm.
\end{abstract}


\vspace{-1em}
\section{Introduction}
\label{sec:intro}
\IEEEPARstart{D}{riven} by the rapid advancement of smart cities and Industry 4.0, high-precision indoor positioning has become a critical foundation for next-generation intelligent systems. Despite the outstanding performance that global navigation satellite systems (GNSSs) have demonstrated in outdoor scenarios, they still face significant challenges indoors\cite{Yang2024}. Against this background, visible light positioning (VLP) has emerged as a promising alternative due to its advantages of high accuracy and low costs\cite{Zhu2024ASurvey,Bastiaens2024}.

VLP technology utilizes light emitting diodes (LEDs) as transmitters to broadcast visible light communication (VLC) signals for positioning. Based on the receiver type, current VLP research can typically be categorized into two main types: photodiode (PD)-based\cite{Huang2020,Xu2024} and camera-based VLP\cite{Hussain2022,VPCA},\cite{VOVLP,He2021,He2024}, each with its own advantages and limitations\cite{Zhu2024ASurvey}. Among them, camera-based VLP extracts visual information by processing captured LED images, and integrates with VLC information for positioning. This allows for high adaptability to environments. Moreover, the wide availability and versatility of cameras make camera-based VLP both practical and easy to deploy\cite{Bastiaens2024}.

Recent advances in camera-based VLP have mainly focused on using the receiver-side camera as the positioning target, with accuracy ranging from centimeters to decimeters. For instance, Hussain \emph{et al.}\cite{Hussain2022} addressed the perspective-$n$-point (P$n$P) problem and proposed a single rectangular LED-based VLP algorithm. To improve the performance of circular LED-based VLP, Zhu \emph{et al.}\cite{VPCA} introduced a perspective circle and arc algorithm to address the duality caused by the pinhole projection model. They further introduced the visual odometry (VO) and developed a single circular LED-based VLP method\cite{VOVLP}, improving the accuracy to centimeter level. However, the above algorithms\cite{Hussain2022,VPCA,VOVLP} require modulated LEDs with specific shapes to be installed at known fixed positions, and cameras to be mounted on targets, which may increase the system deployment costs.

\IEEEpubidadjcol

In fact, in typical indoor scenarios such as factory warehouses and underground parking lots, existing infrastructure, e.g., surveillance cameras, can be directly reused\cite{Hu2023}. This enables an alternative approach, the camera-based passive VLP\cite{He2021,He2024}, which treats LEDs as targets and localizes them using pre-calibrated cameras. Compared to\cite{Hussain2022,VPCA,VOVLP}, passive VLP can achieve simultaneous positioning of multiple targets with lower uncertainty simply by attaching a small unmodulated LED to each target. This is particularly suitable for applications such as warehouse object tracking and monitoring, as it can not only avoid the extensive use of onboard cameras, but also reduce the computational load on the terminal side. He \emph{et al.}\cite{He2021,He2024} have conducted preliminary research on this approach. In\cite{He2021}, the authors developed a dual-camera-based VLP system to achieve millimeter-level accuracy. In\cite{He2024}, they further extended it to a multi-camera scene and proposed a camera layout optimization scheme. However, \cite{He2021} and\cite{He2024} rely solely on the linear method to compute the optimal intersection of projection rays, neglecting the inherent nonlinearity in the projection model, thus being sensitive to image noise.

The primary contribution of this letter is a passive VLP algorithm based on Multi-Camera Joint Optimization (MCJO), which comprises two stages. In the first stage, MCJO constructs projection incident rays from multi-view image observations and solves for target positions using the linear least squares (LLS) method. In the second stage, MCJO formulates a reprojection model and employs nonlinear optimization to minimize the reprojection error, using LLS results as initialization for joint refinement. Simulation results show that MCJO can achieve millimeter-level accuracy, with an improvement of 19\% over the state-of-the-art (SOTA) algorithm in\cite{He2024}. Experimental results further show that MCJO can achieve an average position error of 5.63 mm.

\section{System Model}
\label{sec:sys}

\begin{figure}[t]
\centering
\includegraphics[width=0.43\textwidth]{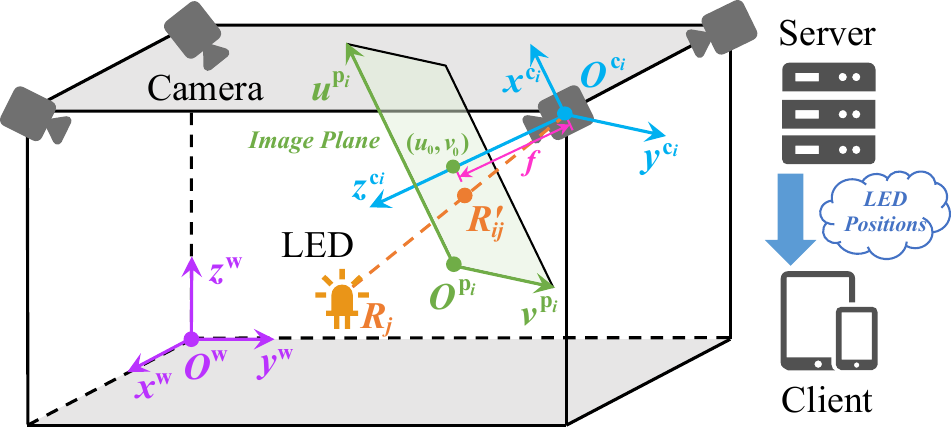}
\caption{The considered passive VLP system.}
\label{fig:scenario}
\vspace{-0.2cm}
\end{figure}

The considered VLP scenario is illustrated in Fig. \ref{fig:scenario}, which is a typical unmodulated passive VLP system\cite{Alijani2025}. In this scenario, each of the $M \geq 1$ positioning targets is equipped with an unmodulated LED point source serving as a transmitter, while $N \geq 2$ cameras are installed on the ceiling as receivers. These cameras continuously capture LED images to acquire visual information, which is then transmitted to a server to simultaneously locate the $M$ targets. Finally, the positions of these LED targets can be fed back to the client terminal.

We adopt the classical pinhole model to describe the imaging process of cameras, and define the following coordinate systems: \rom{1} 3-dimensional (3-D) world coordinate system (WCS) $O^\mathrm{w}\text{--}x^\mathrm{w}y^\mathrm{w}z^\mathrm{w}$, \rom{2} 2-D pixel coordinate system (PCS) $u^\mathrm{p}O^\mathrm{p}v^\mathrm{p}$ on the image plane, and \rom{3} 3-D camera coordinate system (CCS) $O^\mathrm{c}\text{--}x^\mathrm{c}y^\mathrm{c}z^\mathrm{c}$. Since there are $N$ cameras in the scene, there also exist $N$ corresponding CCSs and PCSs, denoted as $\text{CCS}_i$ and $\text{PCS}_i$, respectively. The axes of these CCSs and PCSs are defined such that $x^{\mathrm{c}_i}$-axis aligns with $u^{\mathrm{p}_i}$-axis, $y^{\mathrm{c}_i}$-axis aligns with $v^{\mathrm{p}_i}$-axis, and $z^{\mathrm{c}_i}$-axis is perpendicular to image plane $u^{\mathrm{p}_i}O^{\mathrm{p}_i}v^{\mathrm{p}_i}$. Hereinafter, we use $[\,\cdot\,;\,\cdot\,]$ to denote the vertical stacking of matrices, vectors, or numbers, to form a new matrix or vector. In particular, we denote $[\bm{v};1]$ as $\widetilde{\bm{v}}$, where $\bm{v}$ is a column vector.

In the considered system, the intrinsics of all cameras are the same, which are $\bm{K} \triangleq \left[ \begin{smallmatrix} f_x & 0 & u_0 \\ 0 & f_y & v_0 \\ 0 & 0 & 1 \end{smallmatrix} \right]$, where $f_x$ and $f_y$ are the focal lengths in pixels, and $[u_0;v_0]$ represents the pixel coordinates of the principal point. Since the cameras are installed at fixed positions and orientations, their extrinsics are also known in advance. For a given camera $O^\mathrm{c}$, the transformation from CCS to WCS is represented as:
\begin{equation}
\label{equ:c2w}
    \bm{x}^\mathrm{w} = \bm{R}\cw \bm{x}^\mathrm{c} +\bm{t}\cw,
\end{equation}
where $\bm{x}^\mathrm{w}$ and $\bm{x}^\mathrm{c}$ are arbitrary points in WCS and CCS, respectively; $\bm{R}\cw$ is the rotation matrix representing the camera's orientation; and $\bm{t}\cw$ is the translation vector representing the camera’s position offset. As shown in Fig. \ref{fig:scenario}, an LED target $R$ is captured by camera $O^\mathrm{c}$, and is projected to $R'$ on the image plane. Then, according to the principle of pinhole imaging model, the projection process of point $R$ in CCS onto PCS can be expressed as:
\begin{equation}
\label{equ:c2p}
   z^\mathrm{c}\widetilde{\bm{u}}^\mathrm{p} = \bm{K} \bm{r}^\mathrm{c},
\end{equation}
where $\bm{r}^\mathrm{c}\triangleq[x^\mathrm{c};y^\mathrm{c};z^\mathrm{c}]$ denotes the camera coordinates of $R$, and $\bm{u}^\mathrm{p}\triangleq[u^\mathrm{p};v^\mathrm{p}]$ denotes the pixel coordinates of $R'$.

Based on the above coordinate transformations and projection model, our objective is to estimate the world coordinates $\bm{r}_{j}^\mathrm{w}$ of the $j$-th LED target $R_j$, $1 \leq j \leq M$, from the pixel coordinates $\bm{u}_{j}^{\mathrm{p}_i}$ of the projection points $R_{ij}'$ observed in $\text{PCS}_i$, $1 \leq i \leq N$ on the $i$-th image plane.

\section{Passive VLP Algorithm Based on Multi-Camera Joint Optimization}
\label{sec:alg}

This section details the proposed two-stage MCJO algorithm. The first stage of MCJO is to obtain coarse estimates of target positions based on LLS method, and the second stage refines these estimates by minimizing the total reprojection error through nonlinear optimization.

\subsection{LLS-Based Passive VLP}
\label{ssec:alg1}

Suppose targets $R_j$, $1\leq j\leq M$, are captured by cameras $O^{\mathrm{c}_i}$, $1 \leq i \leq N$. According to Section \ref{sec:sys}, the target $R_j$, its projection $R_{ij}'$ on the image plane $u^{\mathrm{p}_i}O^{\mathrm{p}_i}v^{\mathrm{p}_i}$, and the camera located point $O^{\mathrm{c}_i}$ should lie on the same line. Denote the ray from $O^{\mathrm{c}_i}$ to $R_{ij}'$ as $\ell_{ij}$. For the $N$ cameras, the intersection point of these $N$ rays corresponds to the position of $R_j$.

Let the pixel coordinates of point $R_{ij}'$ be $\bm{u}_{j}^{\mathrm{p}_i}$. Based on \eqref{equ:c2w} and \eqref{equ:c2p}, the normalized direction vector of $\ell_{ij}$ in WCS can be determined as:
\begin{equation}
    \bm{d}^{\mathrm{w}}_{ij} = \bm{R}_{\mathrm{c}_i}^\mathrm{w}\frac{\bm{K}^{-1} \widetilde{\bm{u}}^{\mathrm{p}_i}_{j}}{\big\| \bm{K}^{-1} \widetilde{\bm{u}}^{\mathrm{p}_i}_{j} \big\|}.
\end{equation}
Then, let $\bm{c}_i^\mathrm{w}$ denote the world coordinates of camera $O^{\mathrm{c}_i}$, and the parametric equation of ray $\ell_{ij}$ in WCS can be written as:
\begin{equation}
\label{equ:lij}
    \ell_{ij}^\mathrm{w}:~\bm{x}^\mathrm{w}(\lambda) = \bm{c}_i^\mathrm{w} + \lambda \bm{d}_{ij}^{\mathrm{w}},
\end{equation}
where $\lambda \geq 0$ is a scalar parameter. The target $R_j$ is actually the unique point that minimizes the sum of distances to all $N$ rays. Let $\bm{I}$ denote the identity matrix; then the distance from an arbitrary point $\bm{x}^\mathrm{w}$ in WCS to the ray $\ell_{ij}^\mathrm{w}$ is given by:
\begin{equation}
\label{equ:dij}
    d(\bm{x}^\mathrm{w},\ell_{ij}^\mathrm{w}) = \big\| \big(\bm{I} - \bm{d}^{\mathrm{w}}_{ij} \cdot \big(\bm{d}^{\mathrm{w}}_{ij}\big)^\T\big) \big(\bm{x}^\mathrm{w}-\bm{c}_i^\mathrm{w}\big) \big\|.
\end{equation}

Moreover, we slightly modify the above criterion by instead minimizing the sum of squared distances from the point to all $N$ rays. The point that achieves this minimum is taken as the estimated position $\widehat{\bm{r}}_j^\mathrm{w}$ of the $j$-th target $R_j$:
\begin{equation}
\label{equ:dsquare}
    \widehat{\bm{r}}_j^\mathrm{w} = \arg \min_{\bm{x}^\mathrm{w}} \sum_{i=1}^N \big[d(\bm{x}^\mathrm{w},\ell_{ij}^\mathrm{w})\big]^2.
\end{equation}
The advantage of the formulation in \eqref{equ:dsquare} is that it transforms the nonlinear problem into a linear optimization task. We adopt the LLS method to obtain the closed-form solution to \eqref{equ:dsquare}. In \eqref{equ:dij}, we define:
\begin{subequations}
\begin{numcases}{}
   \bm{A}_{ij} \triangleq \bm{I} - \bm{d}^{\mathrm{w}}_{ij} \cdot \big(\bm{d}^{\mathrm{w}}_{ij}\big)^\T, \\
   \bm{b}_{ij} \triangleq \bm{A}_{ij} \bm{c}_i^\mathrm{w},
\end{numcases}
\end{subequations}
so that \eqref{equ:dsquare} can be further rewritten as:
\begin{equation}
\label{equ:lls1}
    \widehat{\bm{r}}_j^\mathrm{w} = \arg \min_{\bm{x}^\mathrm{w}}\sum_{i=1}^N \big\| \bm{A}_{ij} \bm{x}^\mathrm{w} - \bm{b}_{ij} \big\|^2.
\end{equation}
Then, we define:
\begin{subequations}
\begin{numcases}{}
   \bm{A}_j \triangleq [\bm{A}_{1,j}; \bm{A}_{2,j};\dots;\bm{A}_{N,j}] \in \mathbb{R}^{3N\times3}, \\
   \bm{b}_j \triangleq [\bm{b}_{1,j}; \bm{b}_{2,j};\dots;\bm{b}_{N,j}] \in \mathbb{R}^{3N},
\end{numcases}
\end{subequations}
with which, \eqref{equ:lls1} can be reformulated into the standard form of LLS problem:
\begin{equation}
\label{equ:lls}
    \widehat{\bm{r}}_j^\mathrm{w} = \arg \min_{\bm{x}^\mathrm{w}} \big\| \bm{A}_j\bm{x}^\mathrm{w}-\bm{b}_j \big\|^2.
\end{equation}
Finally, the closed-form solution of \eqref{equ:lls} is obtained by:
\begin{equation}
\label{equ:ini}
    \widehat{\bm{r}}_j^\mathrm{w} = (\bm{A}_j^\T\bm{A}_j)^{-1}\bm{A}_j^\T\bm{b}_j.
\end{equation}

In this way, $\widehat{\bm{r}}_j^\mathrm{w}$ is obtained as the position of the $j$-th target $R_j$. Note that the above process needs to be repeated $M$ times to individually estimate the positions of all $M$ targets. Since this stage only considers a linear model, to improve accuracy, the output results $\widehat{\bm{r}}_j^\mathrm{w}$, $1\leq j\leq M$, will be further refined through nonlinear optimization in Section \ref{ssec:alg2}.

\subsection{Nonlinear Joint Optimization}
\label{ssec:alg2}

In this subsection, we refine the initial positioning results $\widehat{\bm{r}}_j^\mathrm{w}$, $1\leq j\leq M$, obtained in Section \ref{ssec:alg1}. Unlike the process in Section \ref{ssec:alg1} which starts from the receiver-side cameras, we now consider an arbitrary point $\bm{x}_j^{\mathrm{w}}$ in WCS for the $j$-th transmission-side target. According to \eqref{equ:c2w}, if $\bm{x}_j^{\mathrm{w}}$ is captured by camera $O^{\mathrm{c}_i}$, $\bm{x}_j^{\mathrm{w}}$ can be transformed into a point $\bm{x}_j^{\mathrm{c}_i}\triangleq[x_j^{\mathrm{c}_i};y_j^{\mathrm{c}_i}; z_j^{\mathrm{c}_i}]$ in $\text{CCS}_i$. Furthermore, based on \eqref{equ:c2p}, $\bm{x}_j^{\mathrm{c}_i}$ can be projected onto $\text{PCS}_i$ to form a projection point:
\begin{equation}
    \bm{\pi}_i(\bm{x}_j^\mathrm{w}) = \left[ \frac{f_x x_j^{\mathrm{c}_i}}{z_j^{\mathrm{c}_i}}+u_0; \frac{f_y y_j^{\mathrm{c}_i}}{z_j^{\mathrm{c}_i}}+v_0 \right].
\end{equation}

The reprojection pixel error vector $\bm{\epsilon}_i(\bm{x}_j^\mathrm{w})$ between the observed coordinates $\bm{u}_j^{\mathrm{p}_i}$ of the projection point $R_{ij}'$ and the computed projection $\bm{\pi}_i(\bm{x}_j^\mathrm{w})$ is given by:
\begin{equation}
    \bm{\epsilon}_i(\bm{x}_j^\mathrm{w}) = \bm{u}_j^{\mathrm{p}_i}-\bm{\pi}_i(\bm{x}_j^\mathrm{w}).
\end{equation}
Let the output coordinates of target $R_j$ be denoted as $\bm{r}_j^\mathrm{w}$. Since the reprojection error $\|\bm{\epsilon}_i(\bm{x}_j^\mathrm{w})\|$ reflects how well the estimated position aligns with actual image observations, we obtain the optimal coordinates, $\bm{r}_j^\mathrm{w}$, $1\leq j \leq M$, by minimizing the sum of squared reprojection errors across all $N$ image planes for all $M$ targets, i.e.:
\begin{equation}
\label{equ:nlls}
    \{\bm{r}_j^\mathrm{w}\}_{j=1}^M = \mathop{\arg\,\min}_{\substack{\bm{x}_j^\mathrm{w},\forall1 \leq j\leq M}} \sum_{j=1}^M \sum_{i=1}^N \big\| \bm{\epsilon}_i(\bm{x}^\mathrm{w}_j) \big\|^2,
\end{equation}
which is a nonlinear least squares problem. To solve it, we define $\bm{\xi}^\mathrm{w}\triangleq[\bm{x}_1^\mathrm{w};\bm{x}_2^\mathrm{w};\dots;\bm{x}_M^\mathrm{w}] \in \mathbb{R}^{3M}$, and further define the total pixel error vector $\bm{\epsilon}(\bm{\xi}^\mathrm{w})$ as:
\begin{equation}
    \bm{\epsilon}(\bm{\xi}^\mathrm{w})\triangleq[\bm{\epsilon}_1(\bm{x}_1^\mathrm{w});\dots;\bm{\epsilon}_N(\bm{x}_M^\mathrm{w})] \in \mathbb{R}^{2MN},
\end{equation}
so that \eqref{equ:nlls} can be reformulated as: 
\begin{equation}
\label{equ:final}
    \bm{\rho}^\mathrm{w} = \arg \min_{\bm{\xi}^\mathrm{w}} \big\| \bm{\epsilon}(\bm{\xi}^\mathrm{w}) \big\|^2,
\end{equation}
where $\bm{\rho}^\mathrm{w}\triangleq[\bm{r}_1^\mathrm{w};\bm{r}_2^\mathrm{w};\dots;\bm{r}_M^\mathrm{w}] \in \mathbb{R}^{3M}$ denotes the optimal aggregated target vector. Then, \eqref{equ:final} can be solved using the Levenberg-Marquardt (L-M) algorithm\cite{Mouragnon2006}, with $\widehat{\bm{r}}_j^\mathrm{w}$ in \eqref{equ:ini} as the initialization. 

With the optimal target positions $\bm{r}_j^\mathrm{w}$, $1\leq j \leq M$, successfully obtained through the above stage, the system achieves refined and globally consistent results. These positions serve as the final output of MCJO. Note that as analyzed in\cite{He2021}, even in the presence of scene symmetry, MCJO does not suffer from positioning ambiguity under a dual-camera setup.

\subsection{Time Complexity Analysis}
\label{ssec:ca}

\begin{figure}[t]
\centering
\includegraphics[width=0.33\textwidth]{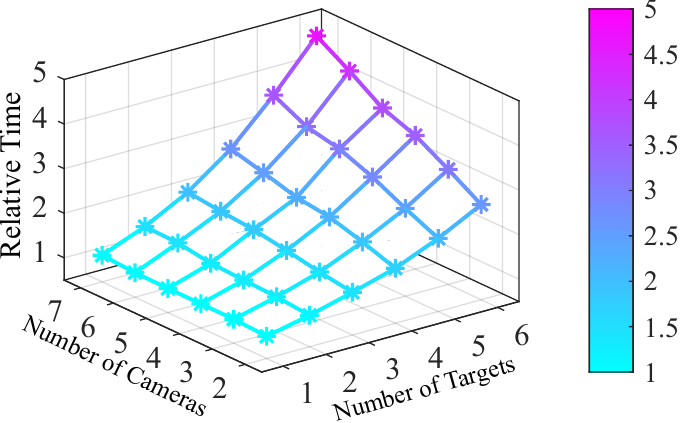}
\caption{RCT versus number of targets and cameras.}
\label{fig:comp}
\vspace{-0.15cm}
\end{figure}

In the first stage, solving \eqref{equ:dsquare} by traversing all $N$ cameras results in a complexity of $\mathcal{O}(N)$, and repeating this process for $M$ targets leads to an overall complexity of $\mathcal{O}(MN)$. In the second stage, the L-M algorithm is used to solve \eqref{equ:final}, and its complexity can be calculated as $I\cdot\mathcal{O}(2MN\cdot(3M)^2+(3M)^3)=\mathcal{O}(18IM^3N+27IM^3)$\cite{Mouragnon2006}, where $I$ denotes the number of iterations. Therefore, the overall complexity of MCJO is $\mathcal{O}(18IM^3N+27IM^3+MN)=\mathcal{O}(IM^3N)$.

We also analyze the relative computation time (RCT) in Fig. \ref{fig:comp} with varying numbers of targets and cameras in simulations. The RCT is defined as the multiple of the time required in the baseline case (2 cameras and 1 target). We can observe that the simulation RCT exhibits a similar trend to the theoretical complexity $\mathcal{O}(IM^3N)$, but with a slower growth rate, indicating high efficiency of MCJO in practice.

\section{Simulations and Experiments}
\label{sec:simu}

In this section, we evaluate the performance of MCJO via simulations and experiments, where the following metrics are used: \rom{1} Position error, defined as the distance between ground-truth and estimated positions of a target; \rom{2} mean/root mean square of position errors (MPE/RMSE) for all targets; and \rom{3} standard deviation (STD) of position errors for all targets. We conduct a multiple-camera-based VLP (MC-VLP) algorithm\cite{He2024} as the SOTA baseline for comparison.

\subsection{Simulation Setup and Results}
\label{ssec:sim}

\begin{table}[t]
\scriptsize
\centering
\caption{Simulation Parameters}
\label{tab:sim_para}
\setlength{\extrarowheight}{1pt}
\setlength{\tabcolsep}{5pt}{
\begin{tabular}{|cc|c|} \hline
\multicolumn{2}{|c|}{\textbf{Parameter}} & \textbf{Value} \\\hline
\multicolumn{1}{|c|}{Platform} & Room size & $8\m\times8\m\times3\m$ \\\hline
\multicolumn{1}{|c|}{\multirow{5}{*}{Camera}} & Positions & $[0;0;3]$, $[8;0;3]$, $[0;8;3]$, $[8;8;3]$ \\\cline{2-3} 
\multicolumn{1}{|c|}{} & Focus target & $[4;4;1.5]$ \\\cline{2-3}
\multicolumn{1}{|c|}{} & Focal length & $f=3.36\mm$ \\\cline{2-3} 
\multicolumn{1}{|c|}{} & Principle point & $[u_0;u_0]=[2080;1560]$ \\\cline{2-3} 
\multicolumn{1}{|c|}{} & Pixel size & $\Delta x=\Delta y=2.24\,\mathrm{\mu m/px}$ \\\hline
\end{tabular}
}
\vspace{-0.1cm}
\end{table}

In the simulations, we assume an indoor environment, i.e., a rectangular room with cameras mounted on the ceiling and directed at the same focus target. The key parameters are listed in Table \ref{tab:sim_para}, unless otherwise specified. The image noise during camera imaging is modeled as a zero-expectation white Gaussian noise with an STD of $\sigma=3\,\mathrm{px}$\cite{He2024}. All statistical results are averaged over $10,000$ independent iterations. In each iteration, $3$ targets are simultaneously localized, and their positions are independently and randomly generated within the room while ensuring visibility to all cameras. 

\begin{figure}[t]
\centering
\begin{subfigure}{0.21\textwidth}
  \includegraphics[height=1.04\textwidth]{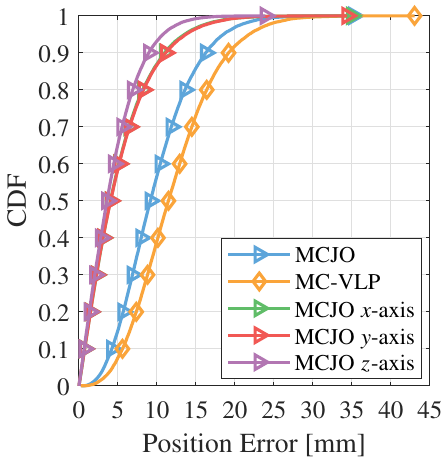}
  \caption{}
  \label{fig:sim_cdf}
\end{subfigure}
\hspace{0.5em}
\begin{subfigure}{0.21\textwidth}
  \includegraphics[height=1.04\textwidth]{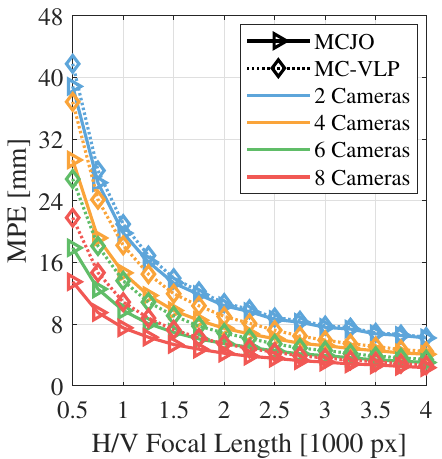}
  \caption{}
  \label{fig:sim_f}
\end{subfigure}

\begin{subfigure}{0.21\textwidth}                       
  \includegraphics[height=1.04\textwidth]{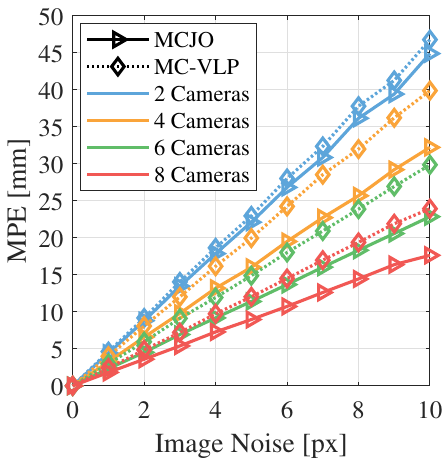}
  \caption{}
  \label{fig:sim_n}
\end{subfigure}
\hspace{0.5em}
\begin{subfigure}{0.21\textwidth}
  \includegraphics[height=1.04\textwidth]{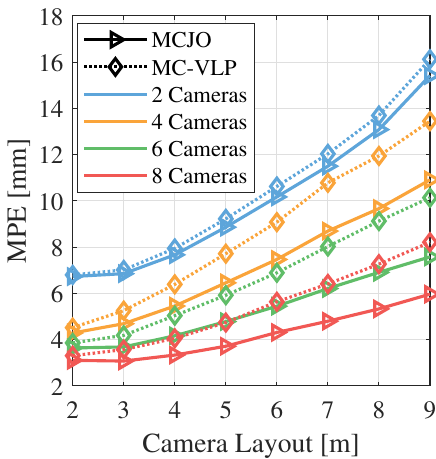}
  \caption{}
  \label{fig:sim_l}
\end{subfigure}
\caption{Simulation Results. (a) CDFs of position errors, (a) MPE versus focal length $f_x=f_y$, (b) MPE versus image noise STD $\sigma$, and (c) MPE versus camera layout distance $L$.}
\label{fig:sims}
\vspace{-0.3cm}
\end{figure}

In Fig. \ref{fig:sims}\subref{fig:sim_cdf}, we compare the performance difference between MCJO and MC-VLP in terms of the cumulative distribution function (CDF) against position error. We can observe that the proposed MCJO algorithm outperforms the baseline, and it can achieve $86$th accuracies of about $15\mm$, while MC-VLP can only achieve $73$th instead. More detailed numerical results are shown in Table \ref{tab:sim_res}, where we can observe the proposed MCJO algorithm consistently outperforms MC-VLP across all evaluation metrics. For instance, MCJO achieves an MPE of $9.69\mm$, representing an improvement of approximately $19\%$ over MC-VLP. In terms of RMSE, MCJO reaches $10.82\mm$, which is about $18\%$ lower than MC-VLP. These results indicate that MCJO can achieve not only higher accuracy but also greater robustness and consistency.

\begin{table}[t]
\scriptsize
\centering
\caption{Simulation Results}
\label{tab:sim_res}
\setlength{\extrarowheight}{1pt}
\setlength{\tabcolsep}{4pt}{
\begin{tabular}{|c|c|c|P{0.85cm}|P{0.85cm}|P{0.85cm}|} \hline
{\footnotesize\textbf{Metrics}} & MCJO & MC-VLP & MCJO \textit{x}-axis & MCJO \textit{y}-axis & MCJO \textit{z}-axis \\\hline
\textbf{MPE [mm]} & \textbf{9.69} & 12.00 & 5.08 & 5.06 & 4.29 \\\hline
\textbf{RMSE [mm]} & \textbf{10.82} & 13.14 & 6.62 & 6.60 & 5.45 \\\hline
\textbf{50\%CDF [mm]} & \textbf{9.08} & 11.36 & 4.02 & 4.07 & 3.59 \\\hline
\textbf{90\%CDF [mm]} & \textbf{16.07} & 19.21 & 10.98 & 11.12 & 8.95 \\\hline
\textbf{STD [mm]} & \textbf{4.74} & 5.34 & 4.25 & 4.24  & 3.36 \\\hline
\end{tabular}
}
\vspace{-0.1cm}
\end{table}

We also evaluate the position error of MCJO along each axis of WCS. As illustrated in Fig. \ref{fig:sims}\subref{fig:sim_cdf} and Table \ref{tab:sim_res}, the errors along $x^\mathrm{w}$- and $y^\mathrm{w}$-axes are nearly identical, while the error along $z^\mathrm{w}$-axis is slightly smaller. This is because all cameras are positioned with a layout distance of $L=8\m$, whereas the room height is only $3\m$. Consequently, the horizontal distances between targets and cameras are generally larger than the vertical distances. As indicated by \eqref{equ:c2p}, targets farther from the camera tend to have a larger $z^\mathrm{c}$, and thus tend to exhibit larger errors. Therefore, the slightly smaller errors along $z^\mathrm{w}$-axis are consistent with the system configuration.

In Figs. \ref{fig:sims}\subref{fig:sim_f}--\subref{fig:sim_l}, we evaluate the impact of focal length, image pixel noise, and camera layout distance on positioning accuracy. As shown in Fig. \ref{fig:sims}\subref{fig:sim_f}, the MPEs of both MCJO and MC-VLP decrease as the focal length increases. This trend aligns with \eqref{equ:c2p}, where a larger focal length reduces the impact of image noise $\delta\bm{u}^\mathrm{p}$ in PCS on the back-projected coordinates $\bm{x}^\mathrm{c}$ in CCS. In Figs. \ref{fig:sims}\subref{fig:sim_n} and \subref{fig:sim_l}, we observe that position errors increase with higher image noise and larger camera layout distances. This is because, according to \eqref{equ:c2p}, a larger noise STD $\sigma$ leads to greater deviations in $\delta\bm{u}^\mathrm{p}$, and a wider layout distance $L$ can increase the $z^\mathrm{c}$ of the target, and both can amplify errors. Furthermore, we also vary the number of cameras to examine its effect on system performance. We can observe that increasing the number of cameras effectively reduces errors, and under the same number of cameras, MCJO consistently achieves higher accuracy. This is because, while only two cameras are theoretically sufficient for positioning, adding more cameras can provide more visual information.

\subsection{Experimental Setup and Results}

\begin{figure}[t]
\centering
\includegraphics[width=0.365\textwidth]{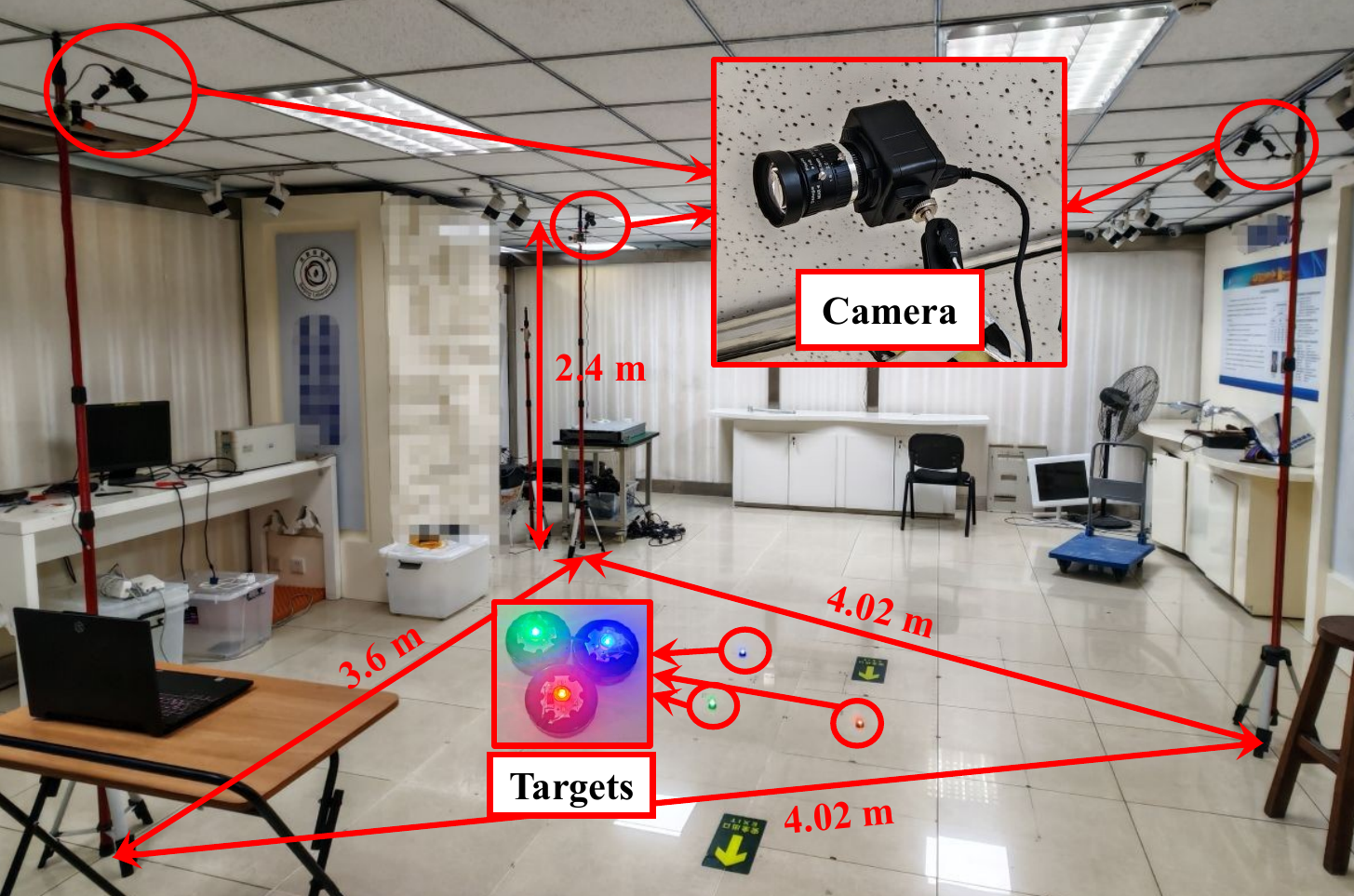}
\caption{The prototype of MCJO positioning system.}
\label{fig:prototype}
\vspace{-0.15cm}
\end{figure}

\begin{table}[t]
\scriptsize
\centering
\caption{Key Experimental Parameters}
\label{tab:exp_para}
\setlength{\extrarowheight}{1pt}
\setlength{\tabcolsep}{5pt}{
\begin{tabular}{|cP{2cm}|P{4cm}|} \hline
\multicolumn{2}{|c|}{\textbf{Parameter}} & \textbf{Value} \\\hline
\multicolumn{1}{|c|}{\multirow{6}{*}{Camera}} & Model & SHL-500W \\\cline{2-3} 
\multicolumn{1}{|c|}{} & Position \& focus target pairs & $[0.05;0.13;2.35]$ \& $[2.0;1.6;0]$, $[3.50;0.09;2.30]$ \& $[1.5;1.4;0]$, $[1.77;3.41;2.26]$ \& $[1.8;1.9;0]$ \\\cline{2-3}
\multicolumn{1}{|c|}{} & Focal length & $f=5\mm$ \\\cline{2-3} 
\multicolumn{1}{|c|}{} & Principle point & $[u_0;u_0]=[1296;972]$ \\\cline{2-3} 
\multicolumn{1}{|c|}{} & Pixel size & $\Delta x=\Delta y=2\,\mathrm{\mu m/px}$ \\\hline
\multicolumn{1}{|c|}{LED} & \multicolumn{2}{c|}{Power $1\,\mathrm{W}$; Half-power angle $120^\circ$; Radius $2.5\mm$} \\\hline
\end{tabular}
}
\vspace{-0.1cm}
\end{table}

To further evaluate the performance of MCJO, we have developed a prototype of MCJO positioning system as shown in Fig. \ref{fig:prototype}, and conducted experiments with key system parameters listed in Table \ref{tab:exp_para}. In the experiments, we use three pre-calibrated cameras directed downward to capture one LED target located within the overlapping field of view (FoV). Note that each camera’s principal axis $z^\mathrm{c}$ is oriented toward a different focus target to ensure a broader overlapping FoV. To ensure the experimental reliability, we calibrate the camera intrinsics using the method in\cite{Zhang2000}, and the extrinsics using the P$n$P algorithm\cite{pnp} with $n = 196$.

\begin{figure*}[t]
\centering
\begin{subfigure}{0.33\textwidth}\centering
  \includegraphics[height=3.5cm]{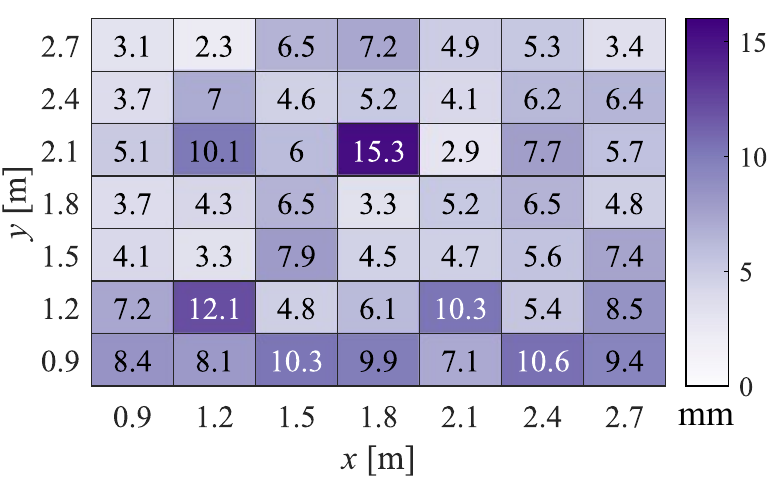}
  \caption{}
  \label{fig:exp1}
\end{subfigure}
\begin{subfigure}{0.33\textwidth}\centering
  \includegraphics[height=3.5cm]{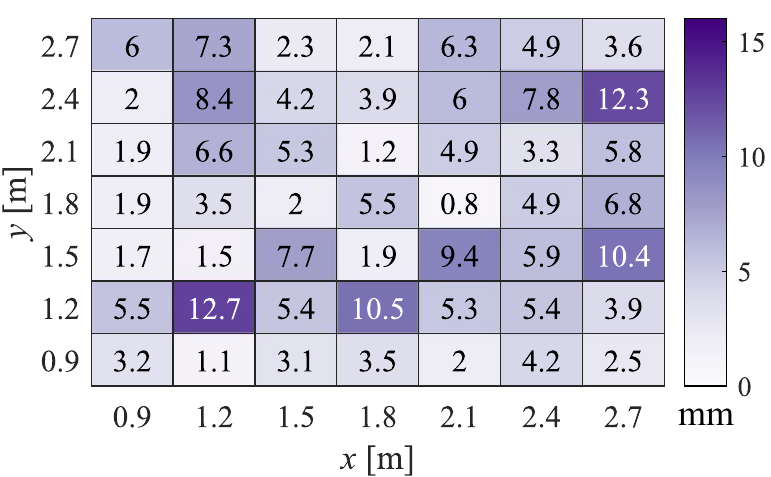}
  \caption{}
  \label{fig:exp2}
\end{subfigure}
\begin{subfigure}{0.29\textwidth}\centering
  \includegraphics[height=3.5cm]{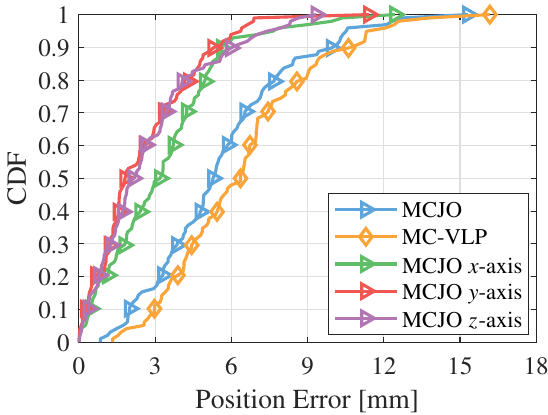}
  \caption{}
  \label{fig:exp_cdf}
\end{subfigure}
\caption{Experimental results. (a) Distribution of 3-D position errors at a height of $0\m$. (b) Distribution of 3-D position errors at a height of $0.3\m$. (c) CDF comparison of 3-D position errors.}
\label{fig:exps}
\vspace{-0.5cm}
\end{figure*}

We collected a total of $98$ positioning results on two planes, $z^{\mathrm{w}} = 0\m$ and $z^{\mathrm{w}} = 0.3\m$, by sampling points at $0.3\m$ intervals within the range $[0.9, 2.7]\times[0.9, 2.7]$. The ground-truth positions were determined based on the precisely known $0.6\m$ grid pattern on the floor and a $0.3\m$-high support stand used to obtain the upper-plane samples. The results of MCJO are shown in Figs. \ref{fig:exps}\subref{fig:exp1} and \subref{fig:exp2}. We can observe that the errors are approximately uniformly distributed across both planes, with STDs of $2.65\mm$ and $2.90\mm$, respectively. It is also worth noting that the MPE at $0\m$ is slightly larger than that at $0.3\m$, which are $6.39\mm$ and $4.87\mm$, respectively. This is mainly because points at higher positions are closer to the cameras. Moreover, we also conducted the MC-VLP algorithm\cite{He2024} under the same conditions. In Fig. \ref{fig:exps}\subref{fig:exp_cdf}, we compare the position error CDFs of MCJO and MC-VLP. The experimental 3-D MPE of MCJO is $5.63\mm$, with 90\% CDF corresponding error less than $1\cm$, while MC-VLP only achieves an MPE of $6.44\mm$. These results demonstrate that the proposed MCJO algorithm can achieve millimeter-level positioning accuracy.

The above experiments focused on a single-target scenario. When localizing multiple targets, we propose using unmodulated LEDs of different distinguishable colors as an effective solution to distinguish between them, as shown in Fig. \ref{fig:prototype}. To eliminate ambient light interference and ensure that all LEDs remain clearly visible, the camera exposure time is set to less than $1\,\mathrm{ms}$\cite{VPCA,VOVLP}. Once the pixel coordinates of the LED spots are extracted, the proposed MCJO can be applied.

Finally, it is also worth noting that the movement of targets does not affect system performance, as the system relies solely on LED images captured by cameras. As discussed in Section \ref{ssec:sim}, with a fixed system configuration, the accuracy depends only on the clarity of LED spots in the images. Therefore, the LED brightness needs to be sufficient for the spots to be clearly identifiable. Furthermore, since the algorithm treats LEDs as points without shape or orientation, the LED orientation does not affect the positioning results, as long as the LED remains visible to at least two cameras.

\section{Conclusion}
\label{sec:con}
This letter has proposed the MCJO algorithm for high-precision indoor VLP. Unlike traditional camera-based algorithms, MCJO treats LEDs as passive targets, allowing multiple pre-calibrated cameras to simultaneously localize these targets at the server. Specifically, in the first stage, MCJO employs the LLS method to estimate target positions from multi-view projection rays; In the second stage, it constructs and minimizes a nonlinear reprojection error model to further refine the initial results. Simulation results show that MCJO can achieve millimeter-level accuracy, with an improvement of 19\% over the SOTA algorithm in\cite{He2024}. Experimental results further show that MCJO achieves an average position error of 5.63 mm. Therefore, the proposed MCJO is promising for applications in future indoor VLP systems.

% \section*{Acknowledgment}
% This should be a simple paragraph before the References to thank those individuals and institutions who have supported your work on this article.

% \balance
\bibliographystyle{IEEEtran}
\bibliography{references}

\vfill

% \end{CJK}
\end{document}